# Elucidating the long-range charge carrier mobility in metal halide perovskite thin films

Jongchul Lim,[a] Maximilian T. Hörantner,[a] Nobuya Sakai,[a] James M. Ball,[a] Suhas Mahesh,[a] Nakita K. Noel,[a] Yen-Hung Lin,[a] Jay B. Patel,[a] David P. McMeekin,[a] Michael B. Johnston,[a] Bernard Wenger[a]* and Henry J. Snaith[a]*

[a]Clarendon Laboratory, University of Oxford, Parks Road, Oxford OX1 3PU, UK

E-mail: bernard.wenger@physics.ox.ac.uk, henry.snaith@physics.ox.ac.uk

**Broader context**

Key material parameters, which affect the usefulness of an absorber material in a solar cell, include the light absorption strength and bandgap, charge carrier lifetime, radiative efficiency and charge carrier mobility. For metal halide perovskites, precisely measuring the light absorption, carrier lifetime and radiative efficiency are relatively straightforward through standard spectroscopic measurements. Quantification of these parameters has been greatly beneficial for rapidly advancing the quality of perovskite thin films and ensuing solar cell efficiency. In contrast, charge carrier mobility, which is defined as the ratio between the drift velocity and the electric field, has proven to be much more challenging to quantify. Since metal halide perovskites contain mobile ions, in addition to mobile electrons and holes, the ions redistribute in an applied electric field, which negates the use of standard methods to quantify mobility, such as time-of-flight, space charge limited current and hall-mobility measurements. The most reliable means to determine charge carrier mobility in perovskite semiconductors has been through non-contact optical pump terahertz probe or transient microwave photo-conductivity measurements. However, these measurements are expected to be sensitive to short-range conductivity in metal halide perovskites, and unlikely to be influenced by longer range scattering events at grain boundaries and charge trapping. They are therefore less relevant to optimisation strategies for photovoltaic devices, where the length scale over which the charges have to travel is on the order of microns. Here we introduce a new methodology for determining long-range charge carrier mobility in perovskite absorber layers, and show how the derived mobility is sensitive to changes in the way that the perovskite absorber layer is processed. We believe this study is the first accurate evaluation of photo-induced long-range lateral mobility of metal halide perovskites, and therefore represents a new handle for future optimisation of perovskite solar cells and optoelectronic devices.

**Abstract**

**Many optoelectronic properties have been reported for lead halide perovskite polycrystalline films. However, ambiguities in the evaluation of these properties remain, especially for long-range lateral charge transport, where ionic conduction can complicate interpretation of data. Here we demonstrate a new technique to measure the long-range charge carrier mobility in such materials. We combine quasi-steady-state photo-conductivity measurements (electrical probe) with photo-induced transmission and reflection measurements (optical probe) to simultaneously evaluate the conductivity and charge carrier density. With this knowledge we determine the lateral mobility to be ~ 2 cm$^2$/Vs for CH$_3$NH$_3$PbI$_3$ (MAPbI$_3$) polycrystalline perovskite films prepared from the acetonitrile/methylamine solvent system. Furthermore, we present significant differences in long-range charge carrier mobilities, from 2.2 to 0.2 cm$^2$/Vs, between films of contemporary perovskite compositions prepared via different fabrication processes, including solution and vapour phase deposition techniques. Arguably, our work provides the first accurate evaluation of the long-range lateral charge carrier mobility in lead halide perovskite films, with charge carrier density in the range typically achieved under photovoltaic operation.**

Over the last few years, metal halide perovskites have been shown to exhibit excellent optoelectronic properties[1–3] as the active layers in photovoltaics (PVs) and light emitting diodes (LEDs), which are typically constructed with vertically layered architectures. Furthermore, excellent long-range charge transport (i.e. long charge carrier diffusion lengths)[4–7] across polycrystalline domains has allowed for the development of transistors[8,9] and back-contacted devices[4,10] using in-plane electrodes.

Amongst many other factors, the photo-induced optoelectronic properties of a material and consequently optoelectronic device performance, are directly influenced by the internal charge carrier density. For any given charge generation rate externally provided by a light source, the change in charge carrier density is determined by the steady state condition when the charge carrier recombination rate equates to the charge generation rate. Depending upon the charge carrier density, recombination is typically governed by a first order monomolecular recombination (Shockley-Read-Hall), a second order bimolecular recombination (conduction band (CB) electrons with valence band (VB) holes), and a third order Auger recombination (two electrons and one hole),[11–13] which occur as the charges traverse grains and grain boundaries, while experiencing trapping/detrapping and scattering.[14] Therefore, studying the photo-induced optoelectronic properties and accurately determining the charge carrier density, particularly under similar charge density regimes to those that occur in the optoelectronic device of interest, could lead to important insights into device operation.

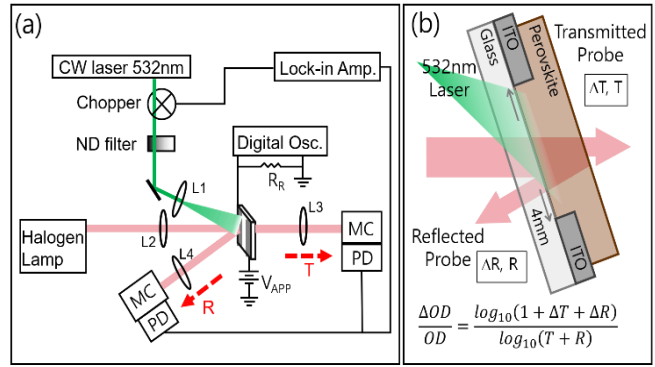

**Fig 1.** Schematic illustration of experimental setup and sample structure. (a) Photo-induced transmission and reflection spectroscopy (PITR) with quasi-steady-state photo-conductivity measurement setup. (b) Sample structure used in this study. Using one sample with this architecture, we measure both PITR and photo-conductivity, with and without external bias, respectively.

Various characterization techniques have been previously employed to investigate the optoelectronic properties of perovskites, such as charge transport and recombination processes, charge transfer across heterojunctions and the fate of photo-excited charge carriers, using either electrical or optical probe measurements. Optical probes include photo-induced absorption (PIA)[3,15,16], transient absorption (TA)[5,17,18] spectroscopy, time resolved microwave conductivity (TRMC)[19,20], optical pump terahertz probe (OPTP) spectroscopy[13,21,22] and time-resolved or quantum yield of photoluminescence (PL)[5,23]. Electrical probes include space charge limited current (SCLC)[17,24], time-of-flight (TOF)[17,25], and hall-mobility measurements[26]. Unfortunately, through these different techniques, the reported values for charge carrier mobility (or diffusion coefficients) vary significantly. The measurements are often in very different charge density regimes, and they also probe conduction over differing length-scales.

Photo-induced absorption (PIA) spectroscopy is a good way to infer the quasi-steady-state photo-induced charge carrier population change at particular pump modulation frequencies and fluences. Hence, it can be used to contribute to a good understanding of the photogenerated charge density changes close to that experienced in the photovoltaic regime ($10^{14}$–$10^{16}$ cm$^{-3}$). A change in free carrier density in any semiconductor, leads to changes in both the imaginary and the real parts of the complex dielectric function.[18] Usually, photo-induced "absorption" (A), is measured by only recording the change in transmittance (T), with the changes in reflectance (R) being neglected. However, since $A = 1 - T - R$, if there is a significant change to the refractive index of the semiconductor during illumination, then the change in reflectivity should not be neglected.[18] With specific relevance to our study here, refractive index changes below the bandgap following photo-excitation, are a result of the free carrier contribution to the dielectric function.[27,28]

Photo-conductivity ($\sigma_{Photo}$) is proportional to the free electron ($N$) and hole ($P$) densities and the average charge carrier mobility ($\mu$), following,

$$\sigma_{Photo} = e\mu(N + P), \quad (1)$$

where $e$ is the elementary charge. Therefore, we propose that by combining quasi-steady-state in-plane $\sigma_{Photo}$ measurements, while estimating the free charge carrier density via photoinduced transmission and reflectance, we can determine the long-range lateral charge carrier mobility.

In ideal semiconductors, the mobility is independent of charge carrier density until carrier-carrier scattering effects become important.[29] In polycrystalline thin films, charge transport is additionally influenced by scattering/trapping processes that can be dependent on charge density, both within the grains and at grain boundaries, which impacts the performance of solar cells. Being able to use $\sigma_{Photo}$ to study the charge carrier density dependent mobility should therefore allow us to better understand losses in performance due to non-ideal long-range charge transport.

Herein we investigate the long-range in-plane $\sigma_{Photo}$, change in charge carrier density and mobility of lead halide perovskite polycrystalline thin films. To do this we have developed a simple and powerful technique for measuring photo-induced transmission and reflection (PITR), the data from which we optically model to determine changes in complex refractive index, and consequently charge carrier density within the perovskite film. We evaluate the charge carrier mobilities within perovskite films fabricated through different processing routes. This allows for an accurate evaluation of the "long-range" lateral mobility in these materials. Our new method has various advantages over other techniques and significantly supports the understanding of photo-induced optoelectronic properties of perovskite materials for various device architectures.

**Results and discussion**

**Photo-induced change in refractive index**

With photo-induced absorption spectroscopy, we can estimate the photo-induced change in charge carrier population within a material at particular pump modulation frequency and fluence. While the photo-induced "absorption" (A) is usually measured by only recording the change in transmittance (T), since $A = 1 - T - R$ in order to properly determine the photo-induced absorption change, it is essential to record both the transmission and reflection spectra. We have therefore modified our experimental set-up, which was originally set up following Ford,[30] to include an additional detector and monochromator to record the reflection signal, as we depict in Fig 1a. During the measurement, the lock-in amplifier generates a specific internal frequency and synchronizes both photodetectors with the frequency of an optical chopper wheel. Periodic square pulse excitation source is realized by chopping a continuous-wave laser with a wavelength of 532nm at 70 Hz and attenuating its intensity with neutral density filters. We detect a periodic change in both transmission and reflection at the same frequency. After sequentially measuring the transmission and reflection changes, we calculate the change in optical density (OD) using,[31]

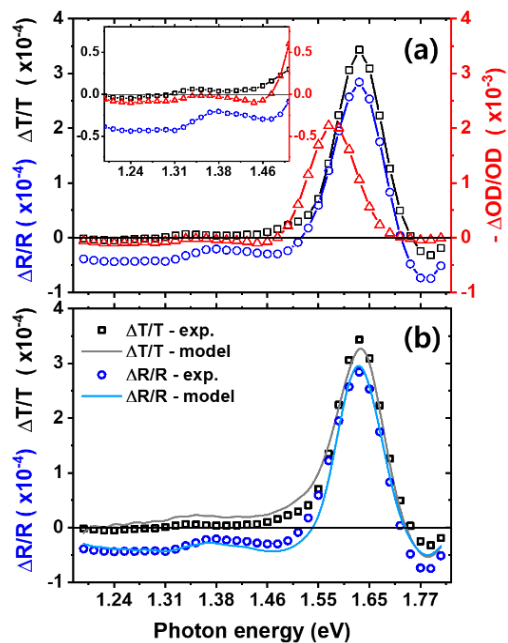

**Fig 2**. Photo-induced transmission and reflection spectra from experiment and fitting. (a) Photo-induced changes in transmission (black squares) and reflection (blue circles) with the calculated optical density change (red triangles, in negative sign for comparison) on the right axis as a function of incident photon energy (excitation density 1.01x10$^{20}$ cm$^{-1}$s$^{-1}$). The inset shows the spectra below the bandgap. (b) Measured photo-induced changes in transmission (black squares) and reflection (blue circles) plotted alongside simulated data (black and blue solid lines, correspondingly) using an optical transfer matrix model using Maclaurin's formula for the Kramers-Kronig transform.

$$\frac{\Delta OD}{OD} = \frac{\log_{10}(1+\Delta T+\Delta R)}{\log_{10}(T+R)}, \qquad (2)$$

where $\Delta OD$, $\Delta T$ and $\Delta R$ are photo-induced changes of optical density, transmittance and reflectance, respectively. The transmitted and reflected light are spectrally resolved and detected through the both monochromators. For quasi-steady-state photo-conductivity (which we simply term, $\sigma_{Photo}$) measurements, we apply a small DC bias voltage to induce an electric field and the digital oscilloscope is used to detect the current flow by measuring the voltage signal across a 1 kΩ resistor, which is in series with the conductivity device and voltage source.

To demonstrate the technique, we prepare polycrystalline MAPbI$_3$ perovskite films using an acetonitrile-based solvent protocol[32] (later referred to as MA-ACN). We spin-coated MAPbI$_3$ films (~ 400 nm) on top of a glass substrate with pre-patterned in-plane indium-tin oxide (ITO) electrodes. This MA-ACN technique leads to ultra-smooth films, minimizing optical scattering effects, resulting in predominantly specular reflection and transmission. We outline additional information on the device preparation in the experimental section. By using this setup together with the in-plane electrode device architecture, which we show in Fig 1b, we can simultaneously perform optical and electrical measurements to gather transmittance (T), reflectance (R), photo-induced changes of transmittance ($\Delta T/T$), reflectance ($\Delta R/R$) and $\sigma_{Photo}$ data, in a single experiment.

In Fig 2a, we plot $-\Delta OD/OD$ to compare the shape of spectra together with the $\Delta T/T$ and $\Delta R/R$. We observe the photobleaching peak at the band edge (around 1.63 eV) in both transmission and reflection.[18,33] This strong change in absorption near the bandgap originates from band filling[34] and bandgap renormalization[18]. However, since our measurement is quasi-steady-state, where most photoexcited charge carriers will have thermalized, we expect band filling to dominate, resulting in the bleaching we observe near the band edge. We note that the peak of the $\Delta OD/OD$ spectra, is at lower energies than the peak of the $\Delta T/T$ and $\Delta R/R$ spectra. This is simply due to the shape of the OD spectrum, changing in an opposite fashion across the band edge, in comparison to the T and R spectra. This shifts the $\Delta OD/OD$ peak to lower energies, since OD is the denominator in this function. We illustrate this in Fig S1.

From Fig 2a, we see that the transmission change in the sub-bandgap region, below 1.4 eV in particular, is small but slightly positive (bleaching). However, we do note that in some previous measurements this has been significant, and most likely erroneously interpreted as a bleaching due the filling of sub-bandgap trap states.[35] Significantly, we observe a more negative $\Delta R/R$ signal in this region, consistent with reduced reflectance. Combining these spectra, we determine a small sub-bandgap increase in optical density ($\Delta OD/OD$ signal).

To explain these transmission and reflection changes, we model and extract the change in complex refractive index $\Delta n(\lambda) + i\Delta k(\lambda)$. We use the transfer matrix model approach, together with the Kramers-Kronig relationship.[36,37] We let the initial complex refractive index of perovskite be $n_0(\lambda) + ik_0(\lambda)$. Upon photo-excitation, the refractive index changes to $n_1(\lambda) + ik_1(\lambda)$. To find $n_1(\lambda) + ik_1(\lambda)$, we searched through the space of possible solutions until we found the $n'(\lambda) + ik'(\lambda)$ that, when fed into an optical transfer matrix model, best reproduces the measured $R_1$ and $T_1$. We chose the trial solutions to be Kramers-Kronig consistent, using a numerical implementation of Maclaurin's formula for the Kramers-Kronig transform.[37] The obtained $n'(\lambda) + ik'(\lambda)$ is equal to $n_1(\lambda) + ik_1(\lambda)$, within the bounds of fitting error. It is then straightforward to calculate $\Delta n(\lambda) + i\Delta k(\lambda) = n_1(\lambda) + ik_1(\lambda) - n_0(\lambda) - ik_0(\lambda)$. We take the initial complex refractive index $n_0(\lambda) + ik_0(\lambda)$ from measurements in the literature.[38] Indeed, we employ global fitting for the full range of each spectrum instead of a specific sub-range (e.g. sub-bandgap) for better reliability. In Fig 2b we show both the experimental and fitted spectra, exhibiting close agreement throughout the entire spectrum. To generate different free carrier densities in the material, we illuminate at varying laser excitation densities, and conduct the fitting process to extract $\Delta n$ using the same protocol.

**Determining the charge carrier density and long-range mobility**

We show the reflectance change of the perovskite after excitation in Fig 3a. We also show on the same graph, our determined change in refractive index, $\Delta n$. The refractive index shows large variations near the band edge. Moreover, a negative change appears below the bandgap, which at low photon energy, increases with reducing photon energy.[27] The carrier induced change in refractive index are related to changes of absorption through Kramers-Kronig relations, and can be explained by two effects in this charge density range: band filling and free carrier absorption.[27] Firstly, after photo-excitation, the bottom of the conduction band minimum (CBM) will be filled by electrons that were excited from the valence band maximum (VBM). Consequently, the minimum energy required to excite an electron across the bandgap is increased giving rise to a widening of the bandgap.[34] This band filling effect results in the strong feature we observe near the band edge. Secondly, free charge carriers can absorb an incident photon and be excited to a higher intraband energy state, which is known as the plasma effect.[27,28,39] This increased free carrier intraband absorption following photo-excitation, results in a corresponding change in the real part of the complex refractive index, which is well described by the Drude model.[27,40] Considering these phenomena, our results are consistent with what is observed in other direct bandgap semiconductors, such as GaAs,[27] where $\Delta n$ shows large changes near the bandgap energy, $E_g$, due to the band filling effect, and approaches zero at photon energies 0.1 to 0.2 eV above or below $E_g$. At energies far below the bandgap, the change in refractive index becomes increasingly negative, as a consequence of the increased free carrier density. We expect the refractive index to change with wavelength ($\lambda$) and free electron ($N$) and hole ($P$) densities, following,[27]

$$\Delta n = -\left(\frac{e^2 \lambda^2}{8\pi^2 c^2 \epsilon_0 n_0}\right)\left(\frac{N}{m_e^*} + \frac{P}{m_h^*}\right), \qquad (3)$$

where $c$ is the speed of light, $\varepsilon_0$ is the permittivity of free space, $n_0$ is the refractive index, and $m_e^*$ and $m_h^*$ are the effective masses of electrons and holes, respectively.

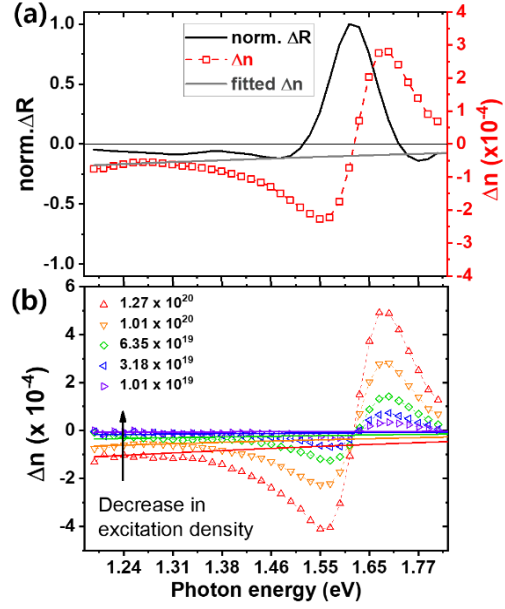

**Fig 3**. Photoinduced change in refractive index and fitting for charge carrier density. (a) Light induced reflection change obtained from experiment, and refractive index change calculated from optical modeling. (b) Excitation density (photons cm$^{-3}$s$^{-1}$) dependent light induced refractive index change (open symbols, dashed line is only to guide the eyes) and fitted data (solid lines) by equation 3.

At energies far enough below the bandgap, the contribution of the band filling effect to $\Delta n$ are negligible. Therefore we fit the lower energy section of the $\Delta n$ spectrum, between 1.2 and 1.3eV, to equation 3, and show such a fit as a light grey line in Fig 3a. Within this energy range, we assume that the contribution to $\Delta n$ from the band filling is negligible, hence $\Delta n$ is predominantly related to the free carrier density. For our fitting we have used equal values of effective masses for electrons and holes from literature ($m_e^* = m_h^* = 0.208\ m_e$)[41], and assumed equal number densities for free electrons and holes.[42,43] To investigate the relationship between refractive index and free carrier density, we repeated the same measurement (Fig S2) and fitting described above, over a range of excitation densities.

With knowledge of the photo-induced charge carrier density, it is possible to determine the average mobility of electrons and holes, which we simply term $\mu$, if the change in $\sigma_{Photo}$ is also known, following the relationship (1). We note that the average mobility, $\mu$, which we present here, is half the value of the sum of electron and hole mobilities, $\Sigma\mu$, usually used for microwave and THz mobility estimations. Therefore, we measure the PITR and $\sigma_{Photo}$ sequentially using the same excitation density and laser illumination for the same device, as we illustrate in the schematic of Fig 1b. For PITR we probe the region between the electrodes, and for $\sigma_{Photo}$, we apply an external bias voltage to the electrodes. The resistance of the perovskite film, reduces under photo-excitation. Therefore, we measure the $\sigma_{Photo}$ by monitoring the voltage drop through a small resistor, which is in series with the in-plane conductivity device, following relationship,[14,44,45]

$$\sigma_{Photo} = \left(\frac{V_R}{R_R(V_{App}-V_R)}\right)\left(\frac{l}{w \times t}\right), \qquad (4)$$

where $V_R$ is the monitored voltage drop through the resistor, $R_R$ is resistor in series with the conductivity device, $V_{App}$ is the externally applied bias voltage, $l$ is the distance between the in-plane electrodes, $w$ is the channel width, and $t$ is the perovskite film thickness.

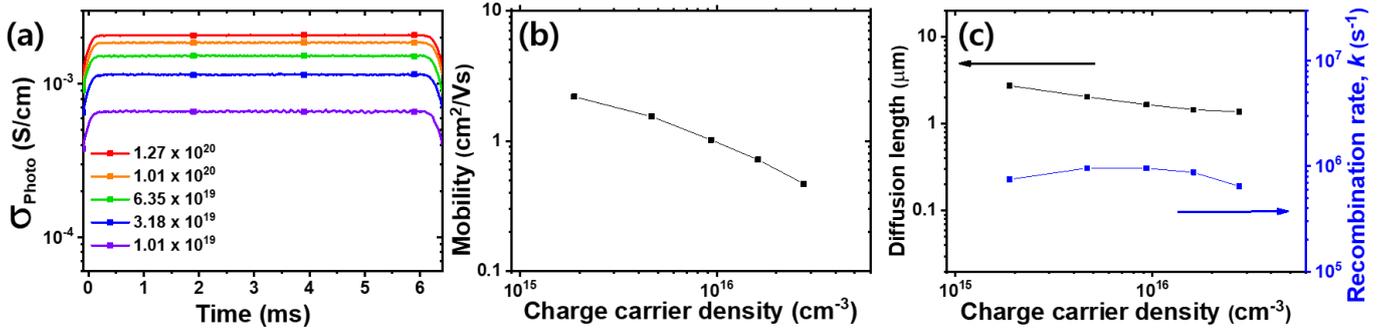

**Fig 4**. Evaluation of photo-induced optoelectronic properties of MAPbI$_3$. (a) Excitation density (photons cm$^{-3}$s$^{-1}$) dependent $\sigma_{Photo}$ of an MA-ACN polycrystalline film with in-plane electrodes, (b) charge carrier mobility (average between electron and hole) calculated using $\sigma_{Photo}$ and charge density determined by the $\Delta n$ (c) charge carrier diffusion length and pseudo-first order recombination ($k$) rate calculated from the mobility and charge density.

We deem that our system is under quasi-steady-state conditions during most of the pulse period (70Hz, 7ms), i.e. the charge carrier generation and recombination rates are equal, as we elaborate upon in Fig S3a. We note that the measured $\sigma_{Photo}$ value is obtained after 3 min illumination (when an equilibrium is reached) to minimize the time dependent photo-doping effect (as shown in Fig S3b)[43]. We plot the $\sigma_{Photo}$ of the MA-ACN films as a function of excitation density in Fig 4a. We show the mobility in Fig 4b, which we determine to be in the range of between 0.47 to 2.2 cm$^2$/Vs, over the excitation range which we study. We note that this mobility is lower than that determined through non-contact THz and microwave conductivity measurements, where values of the sum of electron and hole mobilities ($\mu_e + \mu_h$) are typically determined to be in the range of 20 to 40 cm$^2$/Vs ($\mu_e$ ~ 10 to 20 cm$^2$/Vs). The THz and microwave derived mobility is considered to probe short-range transport for metal halide perovskites, and is hence representative of the highest mobility within a crystalline domain, but is likely to neglect longer range scattering and trapping at grain boundaries and defects within the perovskite film.[12,13,19,20,22,46,47] This difference might stem from the fact that we determined the lateral mobility using a 4 mm distance between the in-plane electrodes. Therefore, our results here include the effect of charge transport dynamics both within grains and across grain boundaries.[46]

Under quasi-steady-state conditions, even though generation and recombination of charge carriers are continuously cycled, the free carrier density is constant (e.g. $dN/dt = 0$), implying that the recombination rate is equivalent to the generation rate $G$. Moreover, in the low charge carrier density regime, the recombination follows a pseudo-first order rate ($k$),[12,13,22] and therefore we can write the quasi-steady-state conditions as $G \sim kN$. Thus, knowing $G$ and $N$, we can estimate an average carrier lifetime $\tau \sim 1/k$ and corresponding pseudo-first order rate, which we show in Fig 4c.

We can now evaluate the charge carrier diffusion length ($L_D$) from our experimentally determined charge carrier mobility ($\mu$) and carrier lifetime ($\tau$) using the following equation:

$$L_D = \left(\frac{\mu k_B T}{e}\tau\right)^{1/2} \qquad (5)$$

where $k_B$ is the Boltzmann constant and $T$ is temperature. In Fig 4c, we show the lateral charge carrier diffusion length for MA-ACN as a function of charge carrier density. This is in the range of 2.73 to 1.37 microns, consistent with the diffusion length estimated by other means.[5,12] We summarize in Table S1, all the optoelectronic parameters which we have obtained from the above evaluation.

**Comparison of the long-range charge carrier mobility within different perovskite films prepared via different deposition methodologies**

The long-range mobility that we have estimated above is for MAPbI$_3$ films processed from the acetonitrile/methylamine mixed solvent system. However, there are many other processing routes and perovskite compositions used in the research community. Here, we assess if the long-range charge carrier mobility of the perovskite layer is strongly influenced by the thin film processing methodology. To investigate this, and to assess the applicability of our technique to a broader set of materials, we prepared perovskite films using different fabrication processes, which are representative of state of the art perovskite absorbers used in the research community; CH$_3$NH$_3$PbI$_3$ films from thermal co-evaporation (termed MA-evap),[48] Cs$_{0.17}$FA$_{0.83}$Pb(I$_{0.9}$Br$_{0.1}$)$_3$ (termed, FACs),[2,49] and Cs$_{0.05}$(FA$_{0.83}$MA$_{0.17}$)$_{0.95}$Pb(I$_{0.9}$Br$_{0.1}$)$_3$ (termed FAMACs)[2,49,50] (See experimental section for the respective deposition procedures).

In Fig S4 we observe a significant difference in the $\sigma_{Photo}$ signal as a function of excitation density, with the FACs sample showing higher $\sigma_{Photo}$ than the FAMACs, with the MA-evap sample showing the lowest $\sigma_{Photo}$. Knowing that the $\sigma_{Photo}$ is determined by the product of mobility and charge carrier density, increased $\sigma_{Photo}$ could be due to either higher carrier density (longer lifetime), or a higher charge carrier mobility, or a combination of the two. Therefore, in order to determine the charge carrier density for these three different samples, we performed the PITR measurement simultaneously to the $\sigma_{Photo}$ measurements, and plot the $\Delta n$ of the three perovskite films as a function of total excitation density in Fig S5. We show charge carrier mobility of the three different samples (in addition to the MA-ACN film) as a function of charge carrier density in Fig 5.

All films exhibit a lower peak mobility as compared to the MA-ACN film, and surprisingly, the evaporated film, which has relatively small grains, exhibits a higher charge carrier mobility, in comparison to the FACs and FAMACs films. Notably however, due to much faster recombination rate (Fig S6) in the evaporated MAPbI$_3$ films, the steady-state charge carrier density is much lower than all the other films we studied. Our estimated mobility of all three perovskites is now in the range of between 1.19 to 0.51, 0.38 to 0.28 and 0.28 to 0.19 cm$^2$/Vs for MA-evap, FACs and FAMACs, respectively. We believe that these now represent good estimation of the long-range mobility within these perovskite films, and should therefore correlate with the relevant mobility for understanding optoelectronic devices where long-range transport occurs. This also implies that significant improvements should be possible in the present state-of-the-art perovskite solar absorbers, if we now focus efforts upon understanding parameters which can be tuned to improve this long-range mobility.

From our data, we observe that our highest mobilities (MA-ACN) are at the lower end, to an order of magnitude lower, than those estimated from early-time optical pump THz probe spectroscopy. This difference may be because here we determine the long-range mobility, which includes conduction between multiple crystalline grains, and fully influenced by the film surfaces[46,47]. Furthermore, we do observe around one order of magnitude difference in mobility between our samples prepared in different ways. This may be due to differences in properties such as perovskite composition, grain sizes,[1,32,51] trapping/detrapping processes, grain-boundary scattering,[52] charging at grain boundaries, crystallinity and crystal orientation. In Figure S7 we show SEM images of the different samples we have studied here for comparison.

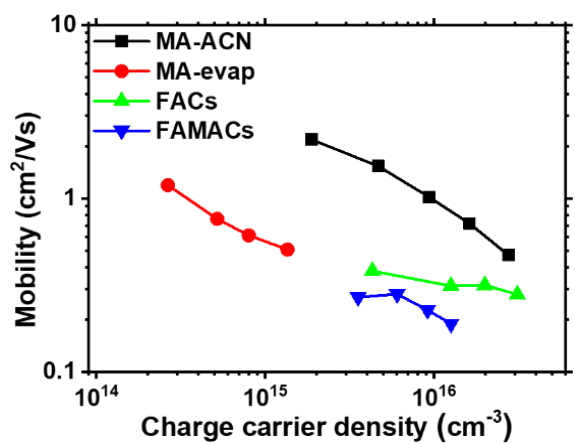

**Fig 5**. Long-range lateral mobility (average electron and hole mobility) with different composition of perovskites obtained by PITR; MA-ACN (MAPbI$_3$ ACN solution processed), MA-evap (MAPbI$_3$ co-evaporation processed), FACs (Cs$_{0.17}$FA$_{0.83}$Pb(I$_{0.9}$Br$_{0.1}$)$_3$ solution processed), FAMACs (Cs$_{0.05}$(FA$_{0.83}$MA$_{0.17}$)$_{0.95}$Pb(I$_{0.9}$Br$_{0.1}$)$_3$ solution processed).

## Conclusions

In conclusion, we have presented a new method for determining the long-range lateral charge carrier mobility within perovskite films by quasi-steady-state optoelectronic techniques. By modeling the photo-induced sub-bandgap changes in the complex refractive index, obtained via photo-induced transmission and reflection (PITR) measurements, we determine the charge carrier density and combine it with the quasi-steady-state photo-conductivity. We evaluate the long-range lateral mobility of $MAPbI_3$, $Cs_{0.17}FA_{0.83}Pb(I_{0.9}Br_{0.1})_3$, $Cs_{0.05}(FA_{0.83}MA_{0.17})_{0.95}Pb(I_{0.9}Br_{0.1})_3$ as a function of free carrier density, and determine mobilities ranging from 2.2 down to 0.2 cm$^2$/Vs depending upon the preparation route and charge carrier density. Despite many years of intense research, to the best of our knowledge, our results represent the first accurate evaluation of the long-range lateral mobility in lead tri-halide perovskite films, and also highlight key advantages of quasi-steady-state method performed under similar excitation densities to a working solar cell. We believe that our technique will help to bring us closer to a complete understanding of long-range charge transport in different metal halide perovskite compounds, and other semiconductors, and gives us a new handle with which to examine and improve perovskite semiconductors.

## Author contributions

J.L. designed and performed all experiments and analyzed the data. M.T.H. and S.M. performed the optical simulations. N.S. conducted SEM measurements. N.S., N.K.N. and Y.H.L. fabricated solution processed perovskite films. J.B.P. and M.B.J. provided ultra-smooth co-evaporated perovskite films. J.M.B. and D.P.M. contributed with discussion and feedback for conductivity measurement. B.W. contributed to the spectroscopic data analysis. J.L. and H.J.S. wrote the manuscript. All authors discussed the results and reviewed the manuscript. B.W. and H.J.S. guided and supervised the overall project.

## Conflicts of interest

There are no conflicts to declare.

## Acknowledgement

This project was funded by EPSRC, Engineering and Physical Sciences Research Council grants, EP/M005143/1 and EP/P006329/1. B.W. acknowledges funding from the European Commission via a Marie-Skłodowska-Curie individual fellowship (REA Grant Number 706552-APPEL).

## Notes and references

# Electronic supplementary Information

**Elucidating the long-range charge carrier mobility in metal halide perovskite thin films**

*Jongchul Lim, Maximilian T. Hörantner, Nobuya Sakai, James M. Ball, Suhas Mahesh, Nakita K. Noel, Yen-Hung Lin, Jay B. Patel, David P. McMeekin, Michael B. Johnston, Bernard Wenger\* and Henry J. Snaith\**

Clarendon Laboratory, University of Oxford, Parks Road, Oxford OX1 3PU, UK

E-mail: bernard.wenger@physics.ox.ac.uk, henry.snaith@physics.ox.ac.uk

**Detail of experiments**

**Figure S1.** Shifted peak of ΔOD/OD spectrum due to shape of the OD spectrum.

**Figure S2.** Photo-induced transmission and reflection changes, and corresponding optical density changes with various excitation densities for MA-ACN perovskite film.

**Figure S3.** Photo-conductivities of MA-ACN perovskite film.

**Table S1.** Photo-induced optoelectronic parameters obtained from the evaluation process as described in main text.

**Figure S4.** Photo-conductivity of three perovskite films prepared in different ways as a function of excitation density.

**Figure S5.** Excitation density dependent light induced refractive index changes from experimental data and fitted data by optical modeling for three perovskite films

**Figure S6.** Photo-induced optoelectronic parameters obtained by the evaluation processes for all perovskite films in our study.

**Figure S7.** Top view images of perovskite polycrystalline films measured by SEM.

**Detail of experiments**

*Preparation of solution processed CH$_3$NH$_3$PbI$_3$ perovskite film*: MAPbI$_3$ (termed MA-ACN) films were fabricated from the ACN/MA compound solvent following a previously published experimental protocol.[1] Briefly, MAI (Dyesol) and PbI$_2$ (TCI Chemicals, 99.5%) were dispersed in ACN (Sigma Aldrich 99.5%, anhydrous) in a 1:1.06 molar ratio, such that the overall concentration of the dispersion was 0.5 M. A solution of methylamine (Sigma Aldrich, 33 wt.% in ethanol) was placed into an ice bath, and using nitrogen as a carrier gas, the solution was degassed of methylamine, which was passed through a drying tube filled with desiccant (CaO and Drierite) before being bubbled into the perovskite dispersion. Methylamine gas was bubbled into the dispersion until all the solid, black crystals were dissolved, leaving a clear, yellow solution. The vial was sealed with a septum cap and the solution kept in the refrigerator at 5 °C until use. This film is stable in air without encapsulation for the duration of the experiment. The film average thickness is 400nm.

*Preparation of Cs$_{0.17}$FA$_{0.83}$Pb(I$_{0.9}$Br$_{0.1}$)$_3$ and Cs$_{0.05}$(FA$_{0.83}$MA$_{0.17}$)$_{0.95}$Pb(I$_{0.9}$Br$_{0.1}$)$_3$ perovskite films*[2–4]: To form the mixed-cation lead mixed anion perovskite precursor solutions, caesium iodide (CsI, Alfa Aesar), formamidinium iodide (FAI, GreatCell Solar), methylammonium iodide (MAI, GreatCell Solar), lead iodide (PbI$_2$, TCI) and lead bromide (PbBr$_2$, Alfa Aesar) were prepared in the way corresponding to the exact stoichiometry for the desired Cs$_{0.05}$(FA$_{0.83}$MA$_{0.17}$)$_{0.95}$Pb(I$_{0.9}$Br$_{0.1}$)$_3$ (termed FAMACs) and Cs$_{0.17}$FA$_{0.83}$Pb(I$_{0.9}$Br$_{0.1}$)$_3$ (termed FACs) compositions in a mixed organic solvent of anhydrous N,N-dimethylformamide (DMF, Sigma-Aldrich) and dimethyl sulfoxide (DMSO, Sigma-Aldrich) at the ratio of DMF : DMSO = 4 : 1. The perovskite precursor concentration used was 1.30 M. The deposition of perovskite layers was carried out using a spin coater in a nitrogen-filled glove box with the following processing parameters: starting at 1000 rpm (ramping time of 4 sec) for 10 sec and then 6000 rpm (ramping time of 6 sec from 1000 rpm) for 35 sec. 10 sec before the end of the spinning process, a solvent-quenching method was used by dropping 300 μL toluene onto the perovskite wet films. The film average thicknesses are 495 and 550nm for FACs and FAMACs, respectively.

***Preparation of co-evaporated CH₃NH₃PbI₃ perovskite film***: CH$_3$NH$_3$PbI$_3$ thin films were deposited on glass substrates using the dual-source thermal evaporation technique as shown previously by Patel *et al.*[5,6] Briefly CH$_3$NH$_3$I and PbI$_2$ were placed in separate crucibles and heated under vacuum (5 x 10$^{-6}$ mbar) until the combined vapour deposition rate was 0.4 Å/s, as measured by the quartz crystal microbalance. Once the rate stabilised, the substrates were exposed to the vapour until the desired thickness of CH$_3$NH$_3$PbI$_3$ was attained. The film average thickness is 400nm.

***PITR setup and measurement***: A 532 nm continuous-wave (CW) laser chopped for 70Hz by internally referenced lock-in amplifier was attenuated by OD filter for various intensity as mentioned in main text. Halogen lamp continuously irradiated on certain masked area of perovskite film for probing the periodic changes of transmission and reflection in photon energy range from 1.20 to 1.85 eV by using long pass and near-infrared (NIR) OD filters. The two identical pre-amplified photodetectors (PDA36a, Thorlabs) are installed on monochromators (Acton SP2300i, Princeton Instruments) for both transmitted and reflected photon detection. Illumination mask is placed on top of sample for focusing pulse laser pump and continuous-wave Halogen lamp probe sources on the same position. The impact of the incidence angle of irradiation for reflections at each interface is fully accounted for in the Transfer Matrix Model. Internal referenced lock-in amplifier synchronizes the measurement of transmission and reflection, for all PL and PIA, then followed by PL subtraction.

***Calculation of Δn and Δk from Reflectance and Transmittance measurements***: We measured the Reflectance $R'$ and Transmittance $T'$ for a photo excited perovskite film. This data was used to calculate the change in refractive index $\Delta n(\lambda) + i\Delta k(\lambda)$ during photo-excitation. The approach is outlined below: Let the initial complex refractive index of perovskite be $n_0(\lambda) + ik_0(\lambda)$. Upon photo-excitation, the refractive index changes to $n_1(\lambda) + ik_1(\lambda)$. To find $n_1(\lambda) + ik_1(\lambda)$, we searched through the space of possible solutions until we found the $n'(\lambda) + ik'(\lambda)$ that, when fed into an optical transfer matrix model, best reproduces the measured $R_1$ and $T_1$. This optimization was done through Powell's Algorithm with a Pythagorean distance function. The trial solutions were always chosen to be Kramers-Kronig consistent, using a numerical implementation of Maclaurin's formula for the Kramers-Kronig transform.[7] The obtained $n'(\lambda) + ik'(\lambda)$ is equal to $n_1(\lambda) + ik_1(\lambda)$, within the bounds of fitting error. It is then straightforward to calculate $\Delta n(\lambda) + i\Delta k(\lambda) = n_1(\lambda) + ik_1(\lambda) - n_0(\lambda) - ik_0(\lambda)$. We take the initial complex refractive index $n_0(\lambda) + ik_0(\lambda)$ from measurements in the literature.[8]

*Quasi-steady-state photo-conductivity*: A continuous diode laser was used at the range of excitation density specified in the main text, with an excitation wavelength of 532 nm and a modulation frequency of 70 Hz. A bias of 25 V was applied across the in-plane electrodes, while the current was monitored by an oscilloscope, as described in detail elsewhere.[9,10] The resistance through the oscilloscope was set by a variable resistor to always be <1% of the sample resistance. We monitored the voltage between the 2 in-plane electrodes through the variable resistor in the circuit to determine the potential dropped across channel length. We then confirmed this by photo-doing effect and frequency effect of the samples at various light intensities, as shown in the Electronic Supplementary Information (**Fig S2**). In both PITR and the photo-conductivity measurement, the square pulse of the laser is generated by a 70 Hz chopper, resulting in a 7 ms periodic laser pulse width, which is much longer than the recombination time of charge carriers.

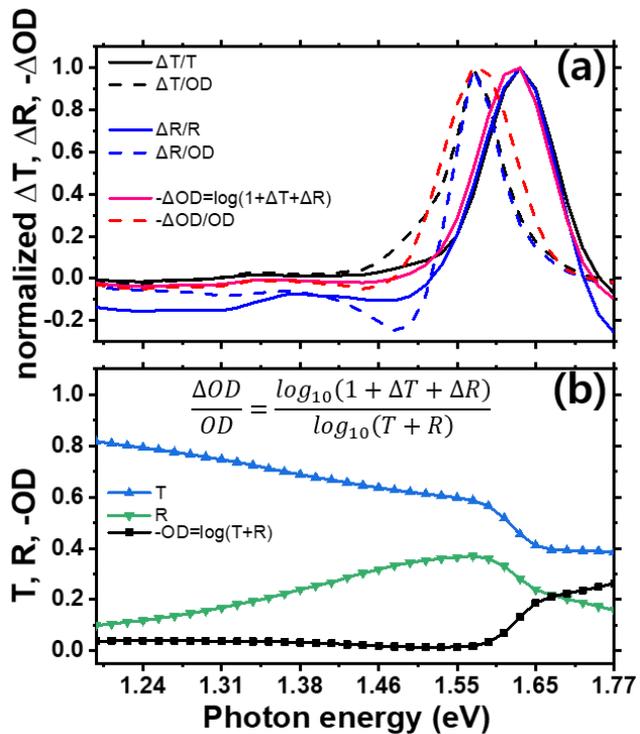

**Figure S1. Shifted peak of ΔOD/OD spectrum due to shape of the OD spectrum (MA-ACN perovskite film).** (a) normalized ΔT/T, ΔR/R and ΔOD spectra, and the normalized spectra of ΔT, ΔR and ΔOD divided by OD for illustrative purposes only. (b) T, R and OD spectra for the calculation of photo-induced changes

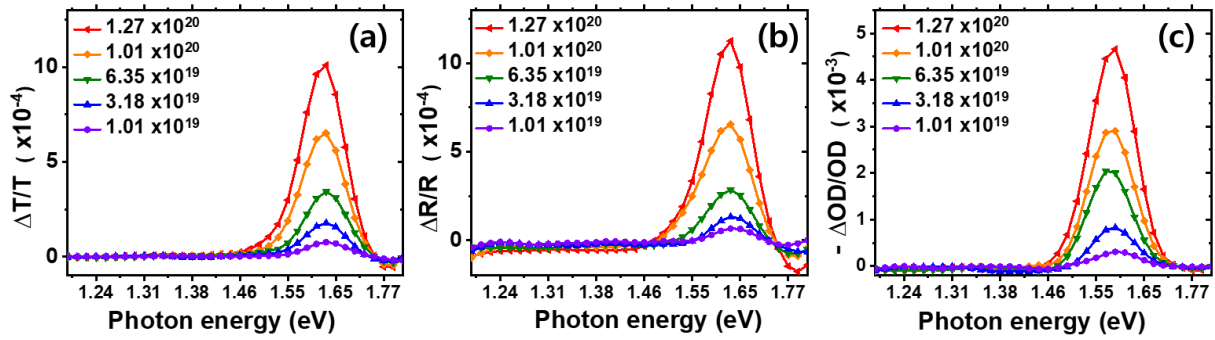

**Figure S2. Photo-induced transmission and reflection changes, and corresponding optical density changes with various excitation densities for MA-ACN perovskite film.** We obtained refractive index changes as a function of excitation density (photons cm$^{-3}$s$^{-1}$) by fitting experimental data ((a) ΔT/T, (b) ΔR/R and (c) ΔOD/OD), as described in experimental section.

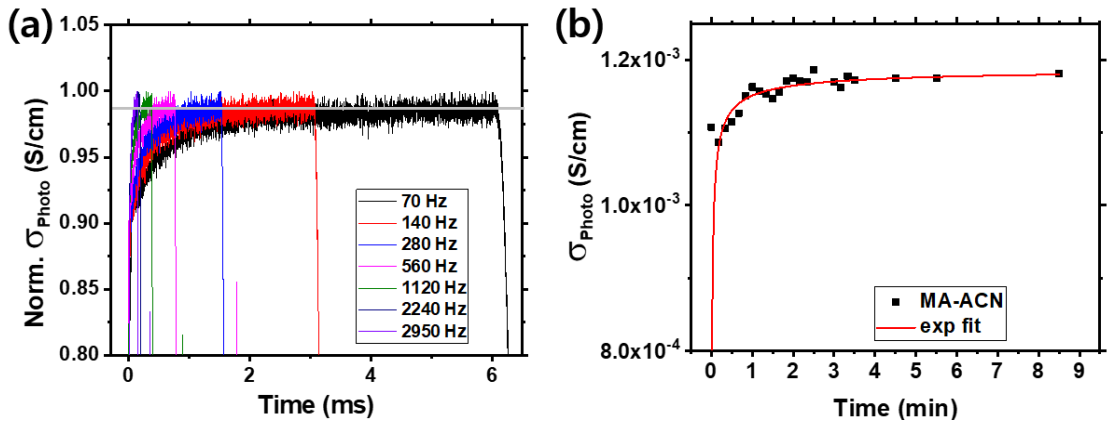

**Figure S3. Photo-conductivities of MA-ACN perovskite film.** (a) Chopping frequency dependent $\sigma_{Photo}$, showing that 70Hz reaches equilibrated state between generation and recombination of charge carrier as a $\sigma_{Photo}$, this indicates that the short pulse illumination time by high chopping frequency is not enough to reach the equilibrium between generation and recombination of charge carriers. (b) Time dependent increase in $\sigma_{Photo}$ as a photo-doping effect.

**Table S1. Photo-induced optoelectronic parameters obtained from the evaluation process as described in main text.** Carrier generation rate ($G$) (cm$^{-3}$s$^{-1}$), charge carrier density (cm$^{-3}$), quasi-steady-state $\sigma_{Photo}$ (S/cm), mobility (cm$^2$/Vs), pseudo-first order recombination rate ($k$) (s$^{-1}$), charge carrier diffusion length (μm), recombination time (μs). MA-ACN: MAPbI$_3$, MA-evap: MAPbI$_3$, FACs: Cs$_{0.17}$FA$_{0.83}$Pb(I$_{0.9}$Br$_{0.1}$)$_3$ FAMACs: Cs$_{0.05}$(FA$_{0.83}$MA$_{0.17}$)$_{0.95}$Pb(I$_{0.9}$Br$_{0.1}$)$_3$.

| | $G$ (cm$^{-3}$s$^{-1}$) | Carrier Density (cm$^{-3}$) | $\sigma_{Photo}$ (S/cm) | Mobility (cm$^2$/Vs) | $k$ (s$^{-1}$) | Diffusion length (μm) | Recombination time (s) |
|---|---|---|---|---|---|---|---|
| | 1.8 x 10$^{22}$ | 2.8 x 10$^{16}$ | 2.1 x 10$^{-3}$ | 0.47 | 6.4 x 10$^5$ | 1.37 | 1.6 x 10$^{-6}$ |
| | 1.4 x 10$^{22}$ | 1.6 x 10$^{16}$ | 1.9 x 10$^{-3}$ | 0.72 | 8.7 x 10$^5$ | 1.45 | 1.2 x 10$^{-6}$ |
| MA-ACN | 8.9 x 10$^{21}$ | 9.3 x 10$^{15}$ | 1.5 x 10$^{-3}$ | 1.02 | 9.5 x 10$^5$ | 1.66 | 1.1 x 10$^{-6}$ |
| | 4.5 x 10$^{21}$ | 4.7 x 10$^{15}$ | 1.2 x 10$^{-3}$ | 1.54 | 9.6 x 10$^5$ | 2.03 | 1.0 x 10$^{-6}$ |
| | 1.4 x 10$^{21}$ | 1.9 x 10$^{15}$ | 6.6 x 10$^{-4}$ | 2.19 | 7.5 x 10$^5$ | 2.74 | 1.3 x 10$^{-6}$ |
| | 4.0 x 10$^{22}$ | 1.4 x 10$^{15}$ | 1.1 x 10$^{-4}$ | 0.51 | 2.9 x 10$^7$ | 0.21 | 3.4 x 10$^{-8}$ |
| MA-evap | 2.8 x 10$^{22}$ | 8.0 x 10$^{14}$ | 7.8 x 10$^{-5}$ | 0.61 | 3.5 x 10$^7$ | 0.21 | 2.8 x 10$^{-8}$ |
| | 1.8 x 10$^{22}$ | 5.2 x 10$^{14}$ | 6.3 x 10$^{-5}$ | 0.76 | 3.4 x 10$^7$ | 0.24 | 2.9 x 10$^{-8}$ |
| | 8.9 x 10$^{21}$ | 2.7 x 10$^{14}$ | 5.1 x 10$^{-5}$ | 1.19 | 3.4 x 10$^7$ | 0.30 | 3.0 x 10$^{-8}$ |
| | 2.3 x 10$^{22}$ | 3.1 x 10$^{16}$ | 1.4 x 10$^{-3}$ | 0.28 | 7.4 x 10$^5$ | 0.99 | 1.4 x 10$^{-6}$ |
| FACs | 1.5 x 10$^{22}$ | 2.0 x 10$^{16}$ | 1.0 x 10$^{-3}$ | 0.32 | 7.3 x 10$^5$ | 1.06 | 1.4 x 10$^{-6}$ |
| | 7.3 x 10$^{21}$ | 1.3 x 10$^{16}$ | 6.3 x 10$^{-4}$ | 0.31 | 5.8 x 10$^5$ | 1.18 | 1.7 x 10$^{-6}$ |
| | 5.8 x 10$^{21}$ | 4.3 x 10$^{15}$ | 2.6 x 10$^{-4}$ | 0.38 | 1.3 x 10$^6$ | 0.86 | 7.5 x 10$^{-7}$ |
| | 2.0 x 10$^{22}$ | 1.3 x 10$^{16}$ | 3.8 x 10$^{-4}$ | 0.19 | 1.6 x 10$^6$ | 0.55 | 6.2 x 10$^{-7}$ |
| FAMACs | 1.3 x 10$^{22}$ | 9.1 x 10$^{15}$ | 3.3 x 10$^{-4}$ | 0.23 | 1.4 x 10$^6$ | 0.65 | 7.1 x 10$^{-7}$ |
| | 6.4 x 10$^{21}$ | 6.0 x 10$^{15}$ | 2.7 x 10$^{-4}$ | 0.28 | 1.1 x 10$^6$ | 0.82 | 9.4 x 10$^{-7}$ |
| | 5.1 x 10$^{21}$ | 3.5 x 10$^{15}$ | 1.5 x 10$^{-4}$ | 0.27 | 1.4 x 10$^6$ | 0.69 | 7.0 x 10$^{-7}$ |

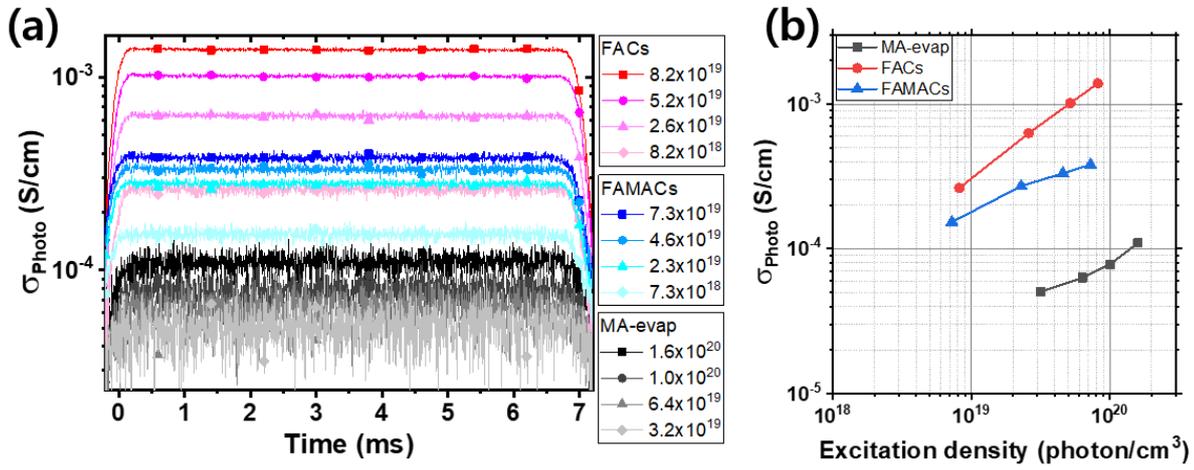

**Figure S4. Photo-conductivity of three perovskite films prepared in different ways as a function of excitation density.** (a) Photo-conductivity under the 7ms light illumination (70Hz modulation, excitation density: photons cm$^{-3}$s$^{-1}$) and (b) the corresponding photo-conductivity values as a function of excitation density (photons cm$^{-3}$s$^{-1}$). MA-evap (MAPbI$_3$ evaporation processed), FACs (Cs$_{0.17}$FA$_{0.83}$Pb(I$_{0.9}$Br$_{0.1}$)$_3$ solution processed) and FAMACs (Cs$_{0.05}$(FA$_{0.83}$MA$_{0.17}$)$_{0.95}$Pb(I$_{0.9}$Br$_{0.1}$)$_3$ solution processed).

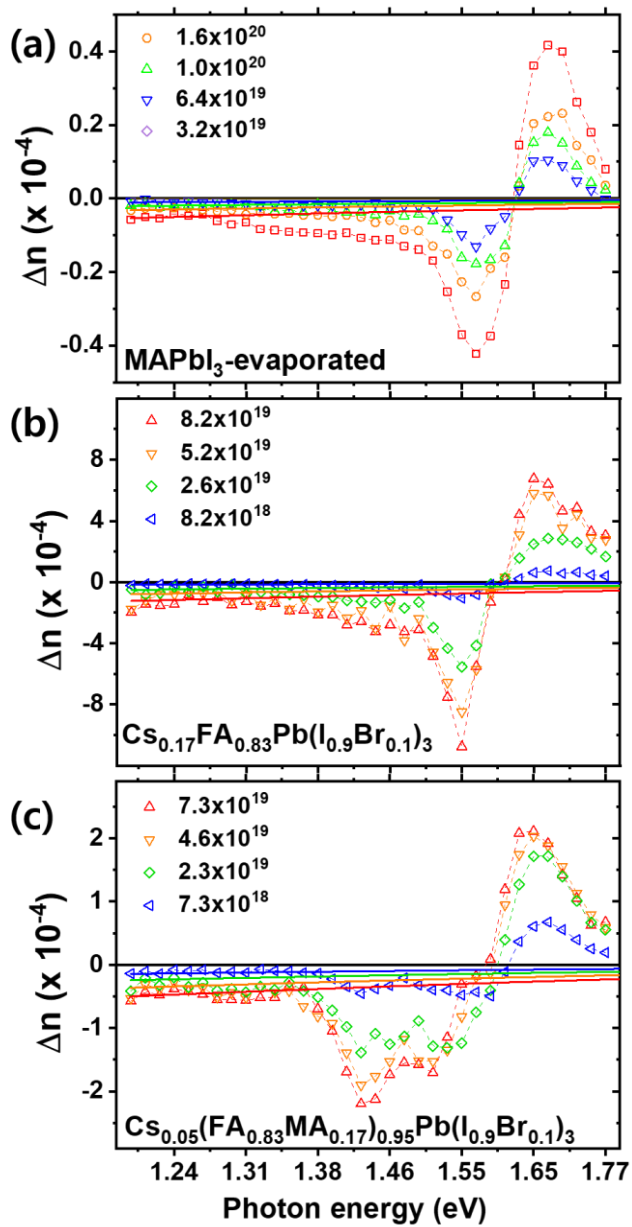

**Figure S5.** Excitation density (photons cm$^{-3}$s$^{-1}$) dependent light induced refractive index changes from experimental data (open symbol, dashed line is only to guide the eyes) and fitted data (solid lines) by optical modeling for three perovskite films; (a) MA-evap (MAPbI$_3$ evaporation processed) (b) FACs (Cs$_{0.17}$FA$_{0.83}$Pb(I$_{0.9}$Br$_{0.1}$)$_3$ solution processed), FAMACs (Cs$_{0.05}$(FA$_{0.83}$MA$_{0.17}$)$_{0.95}$Pb(I$_{0.9}$Br$_{0.1}$)$_3$ solution processed).

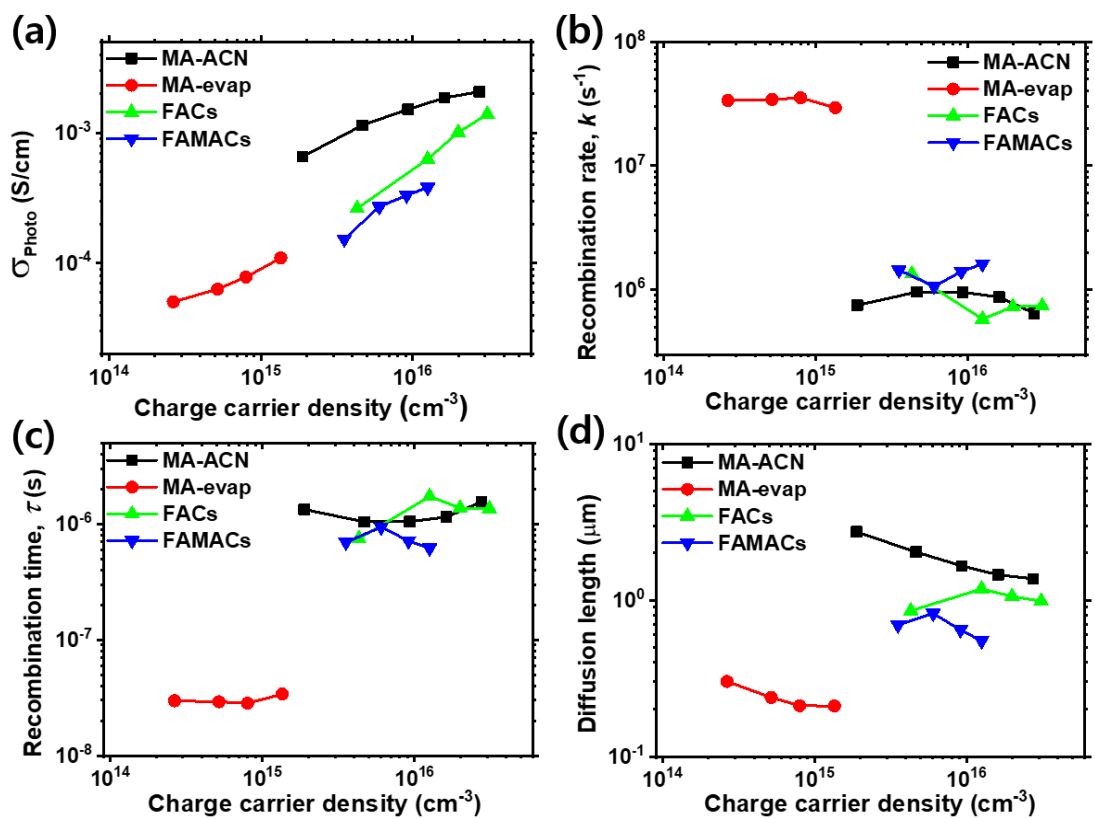

**Figure S6. Photo-induced optoelectronic parameters obtained by the evaluation processes for all perovskite films in our study.** (a) Photo-conductivity (S/cm), (b) pseudo-first order recombination rate (s$^{-1}$), (c) recombination time (s) and (d) charge carrier diffusion length (μm) as a function of charge carrier density (cm$^{-3}$) for MA-ACN (MAPbI$_3$-ACN solution processed), MA-evap (MAPbI$_3$ evaporation processed), FACs (Cs$_{0.17}$FA$_{0.83}$Pb(I$_{0.9}$Br$_{0.1}$)$_3$ solution processed) and FAMACs (Cs$_{0.05}$(FA$_{0.83}$MA$_{0.17}$)$_{0.95}$Pb(I$_{0.9}$Br$_{0.1}$)$_3$ solution processed) films.

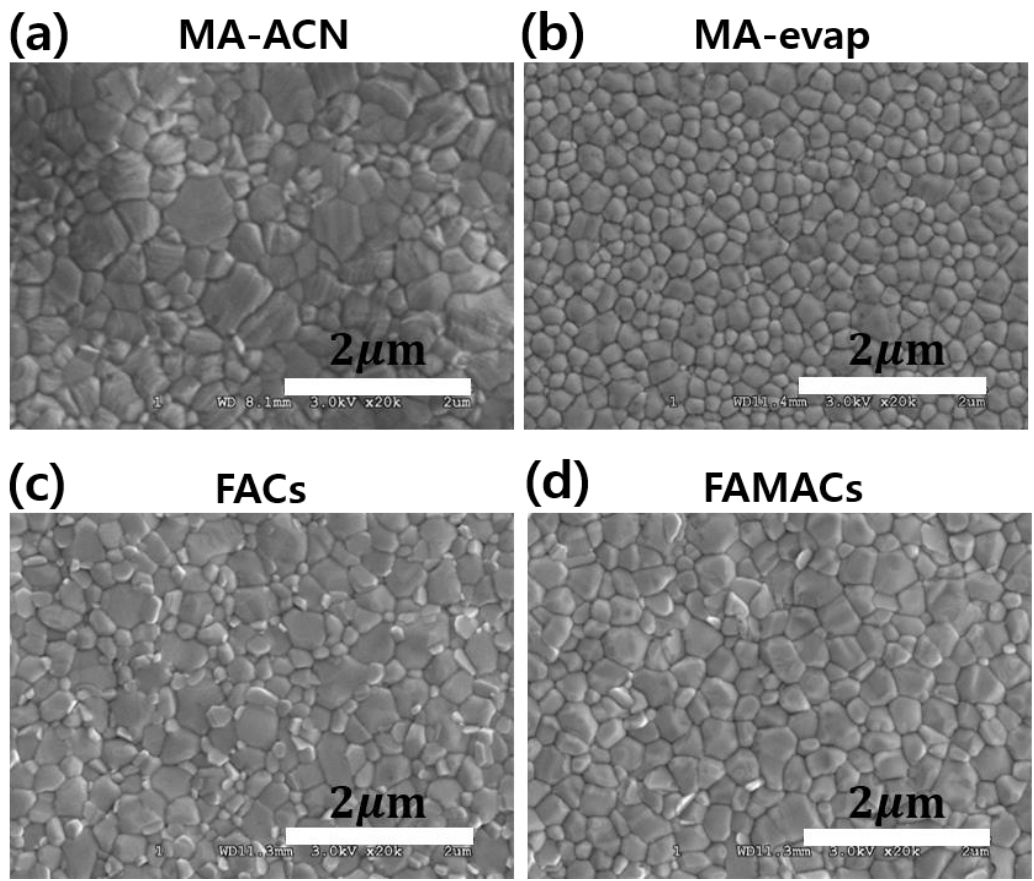

**Figure S7. Top view images of perovskite polycrystalline films measured by SEM.** Sizes of individual domains ranging (a) between 500 nm and 700 nm for MA-ACN film ($MAPbI_3$-ACN solution processed), (b) between 162 nm and 360 nm for MA-evap film ($MAPbI_3$-evaporated), (c) between 236 nm and 585 nm for FACs film ($Cs_{0.17}FA_{0.83}Pb(I_{0.9}Br_{0.1})_3$ solution processed) and (d) between 183 nm and 442 nm for FAMACs film ($Cs_{0.05}(FA_{0.83}MA_{0.17})_{0.95}Pb(I_{0.9}Br_{0.1})_3$ solution processed.